\documentclass[aps,10pt,twocolumn,showpacs]{revtex4}
\usepackage{epsfig}
\usepackage{graphicx}
\begin{document}

% You should use BibTeX and revtex.bst for references
%\bibliographystyle{revtex}

% Use the \preprint command to place your local institutional report
% number  and your conference paper identification number on the
% title page in preprint mode. Multiple \preprint commands are allowed.
%\preprint{\Large\bf {BooNE-TN-127}}
%Title of paper
\title{Boosted Decision Trees as an Alternative to Artificial Neural Networks
for Particle Identification}

% Optional argument for running titles on pages
%\title[]{}

% repeat the \author .. \affiliation  etc. as needed
% \email, \thanks, \homepage, \altaffiliation all apply to the current
% author. Explanatory text should go in the []'s, actual e-mail
% address or url should go in the {}'s for \email and \homepage.
% Please use the appropriate macro for the type of information

% \affiliation command applies to all authors since the last
% \affiliation command. The \affiliation command should follow the
% other information

\author{Byron P. Roe, Hai-Jun Yang}
\email[Corresponding author, e-mail address: ]{yhj@umich.edu}
\affiliation{Department of Physics, University of Michigan, Ann Arbor, MI 48109, USA}
\author{Ji Zhu}
\affiliation{Department of Statistics, University of Michigan, Ann Arbor, MI 48109, USA}
\author{Yong Liu, Ion Stancu}
\affiliation{Department of Physics and Astronomy, University of Alabama, Tuscaloosa, AL 35487, USA}
\author{Gordon McGregor}
\affiliation{Los Alamos National Laboratory, Los Alamos, NM 87545, USA}

\date{\today}

%\homepage[]{Your web page}
%\thanks{}

\begin{abstract}
% insert abstract here

The efficacy of particle identification is compared using
artificial neutral networks and boosted decision trees.
The comparison is performed in the context of the MiniBooNE,
an experiment at Fermilab searching for neutrino oscillations.
Based on studies of Monte Carlo samples of simulated data,
particle identification with boosting algorithms has better
performance than that with artificial neural networks for
the MiniBooNE experiment.
Although the tests in this paper were for one experiment,
it is expected that boosting algorithms will find wide 
application in physics.

\end{abstract}
% insert suggested PACS numbers in braces on next line

%\maketitle must follow title, authors, abstract and \pacs
\maketitle{}

\section{Introduction}

The artificial neural network (ANN) technique has been widely used in data analysis
of High Energy Physics experiments in the last decade. The use of the ANN 
technique usually gives better results than the traditional simple-cut techniques.
In this paper, another data classification technique, {\em boosting}, is introduced 
for data analysis in the MiniBooNE experiment\cite{boone} at Fermi National 
Accelerator Laboratory.  The MiniBooNE experiment is designed to 
confirm or refute the evidence for $\nu_\mu \rightarrow \nu_e$ oscillations at 
$\Delta m^2 \simeq 1 ~eV^2/c^4$ found by the LSND experiment\cite{lsnd}. 
It is a crucial experiment which will imply new physics beyond the standard model
if the LSND signal is confirmed.
Based on our studies, particle identification
(PID) with the boosting algorithm is 20 to 80\% better than that with 
our standard ANN PID technique, the boosting performance relative to that of ANN
depends on the Monte Carlo samples and PID variables.
Although the boosting algorithm was tested in only one experiment, it's anticipated
to have wide application in physics, especially in data analysis of particle 
physics experiments for signal and background events separation.

The boosting algorithm is one of the most powerful learning techniques introduced
during the past decade. The boosting algorithm is a
procedure that combines many ``weak'' classifiers to achieve a final powerful classifier.
Boosting can be applied to any classification method. In this paper, it is applied
to decision trees.
Two boosting algorithms, AdaBoost\cite{adaboost} and $\epsilon$-Boost\cite{eboost},
are considered. A brief description of boosting algorithms is given
in the next section. Our results are presented in Section III, while we summarize 
our conclusions in Section IV.

\section{Brief Description of Boosting}

\subsection{Decision Tree}  
Suppose one is trying to divide events into signal and background and suppose Monte Carlo 
samples of each are available. Divide each Monte Carlo sample into two parts.  
The first part, the training sample, will be used to train the decision tree, 
and the second part, the test sample, to test the final classifier after 
training.  

For each event, suppose there are a number of 
PID variables useful for distinguishing between signal and background.  
Firstly, for each PID variable, order the events by the value of the variable.  
Then pick variable one and for each event value see what happens 
if the training sample is split into
two parts, left and right, depending on the value of that variable.  Pick
the splitting value which gives the best separation into one side having 
mostly signal and the other mostly background.  Then repeat this for each
variable in turn.  Select the variable and splitting value which gives the
best separation.  Initially there was a sample of events at a ``node''.  
Now there are two samples called ``branches''.  For each branch,
repeat the process, i.e., again try each value of each variable for the
events within that branch to find the best variable and splitting point
for that branch.  One keeps splitting until a given number of final branches,
called leaves, are obtained, or until each leaf is pure signal or pure background, 
or has too few events to continue.  This description is a little oversimplified.  
In fact at each stage one picks as the next branch to split, the branch which 
will give the best increase in the quality of the separation.
A schematic of a decision tree is shown in Fig.1, in which 3 variables are
used for signal/background separation: event hit multiplicity, energy,
and reconstructed radial position.

What criterion is used to define the quality of separation between signal
and background in the split?  Imagine the events are weighted with each event
having weight $W_i$.  Define the purity of the sample in a branch
by 
$$P = {\sum_sW_s\over \sum_sW_s +\sum_bW_b},$$
where $\sum_s$ is the sum over signal events and $\sum_b$ is the sum
over background events.
Note that $P(1-P)$ is 0 if the sample is pure signal or pure background.
For a given branch let
$$Gini = (\sum_{i=1}^n W_i)P(1-P),$$
where $n$ is the number of events on that branch.
The criterion chosen is to minimize 
$$Gini_{left\ son} + Gini_{right\ son}.$$

To determine the increase in quality when a
node is split into two branches, one maximizes
$$ Criterion = Gini_{father} - Gini_{left\ son} - Gini_{right\ son}
.$$
 
At the end, if a leaf has purity greater than 1/2 (or whatever is set), then
it is called a signal leaf and if the purity is less than 1/2, it is a
background leaf.  Events are classified signal if they land on a signal 
leaf and background if they land on a background leaf.  The resulting
tree is a {\it decision tree}.

Decision trees have been available for some time\cite{breiman}.  
They are known to be powerful but unstable, i.e., a small change in the 
training sample can give a large change in the tree and the results.

There are three major measures of node impurity used in practice:
misclassification error, the gini index and the cross-entropy.  If we
define p as the proportion of the signal in a node, then the three
measures are:
1 - max(p, 1-p) for the misclassification error,
2p(1-p) for the gini index and 
-plog(p) - (1-p)log(1-p) for the cross-entropy.
The three measures are similar, but the gini index and the cross-entropy
are differentiable, and hence more amenable to numerical optimization.  In
addition, the gini index and the cross-entropy are more sensitive to
change in the node probabilities than the misclassification error.  The
gini index and the cross-entropy are similar.

\begin{figure}
\epsfig{figure=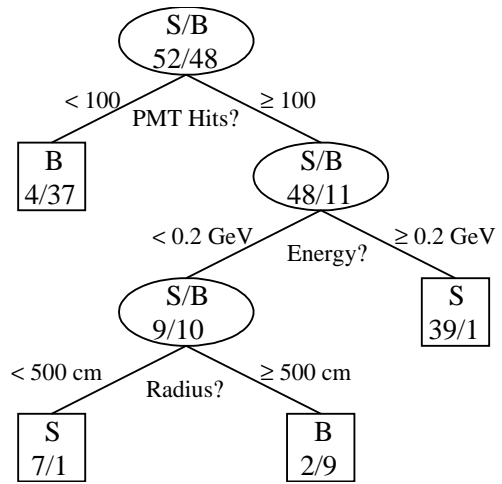,width=7cm,angle=0}
\caption{Schematic of a decision tree. S for signal, B for background.
Terminal nodes(called leaves) are shown in boxes. If signal events are
dominant in one leave, then this leave is signal leave; otherwise,
background leave.}
\end{figure}

\subsection{Boosting}
Within the last few years a great improvement has been 
made\cite{schapire,freund,friedman}.  Start
with unweighted events and build a tree as above.  If a training event is
misclassified, i.e, a signal event lands on a background leaf or a 
background event lands on a signal leaf, then the weight of that event
is increased (boosted). 

A second tree is built using the new weights, no longer equal.  Again
misclassified events have their weights boosted and the procedure is
repeated.  Typically, one may build 1000 or 2000 trees this way.

A score is now assigned to an event as follows.  The event is followed
through each tree in turn.  If it lands on a signal leaf it is given a
score of 1 and if it lands on a background leaf it is given a score 
of -1.  The renormalized sum of all the scores, possibly weighted, is the final
score of the event.  High scores mean the event is most likely signal and
low scores that it is most likely background.  By choosing a particular
value of the score on which to cut, one can select a desired fraction of the
signal or a desired ratio of signal to background.  For those familiar with
ANNs, the use of this score is the same as the 
use of the ANN value for a given event.  For the MiniBooNE experiment,
boosting has been found to be superior to ANNs. 
Statisticians and computer scientists have found that this 
method of classification is very 
efficient and robust.  Furthermore, the amount of tuning
needed is rather modest compared with ANNs.  It works well with
many PID variables.  
If one makes a monotonic transformation of a variable, so 
that if $x_1>x_2$ then $f(x_1)>f(x_2)$, the boosting method gives 
exactly the same results.  It depends only on the ordering according to
the variable, not on the value of the variable.

In articles on boosting within the statistics and computer science 
communities, it is often
recommended that short trees with eight leaves or so be used.  For the
MiniBooNE Monte Carlo samples it was found that large trees with 45 leaves
worked significantly better.  

\subsection{Some Boosting Algorithms}
If there are $N$ total
events in the sample, the weight of each event is initially taken as
$1/N$. Suppose that there are $N_{tree}$ trees and $m$ is the index of an 
individual tree.  Let 
\begin{itemize}
\item $x_i=$  the set of PID variables for the $i$th event.
\item $y_i =1$ if the $i$th event is a signal event
and $y_i=-1$ if the event is a background event.
\item $w_i=$ the weight of the $i$th event.  
\item $T_m(x_i)=1$ if the set of variables for the $i$th event lands that
event on a signal leaf and  $T_m(x_i)=-1$ if the set of variables for that
event lands it on a background leaf.
\item $I(y_i \ne T_m(x_i)) = 1$ if $y_i \ne T_m(x_i)$ and 0 if 
  $y_i = T_m(x_i)$.
\end{itemize}
There are at least two commonly used methods for boosting the weights of the 
misclassified events in the training sample.

The first boosting method is called AdaBoost\cite{adaboost}.
Define for the $m$th tree:
$$err_m ={\sum_{i=1}^Nw_iI(y_i \ne T_m(x_i))\over \sum_{i=1}^Nw_i}. $$
%Calculate:
$$\alpha_m = \beta\times\ln((1-err_m)/err_m).$$
$\beta=1$ is the value used in the standard AdaBoost method.  For the
MiniBooNE Monte Carlo samples, $\beta = 0.5$ has been found to give better results.
Change the weight of each event $i$, $i = 1,...,N$: 
$$w_i\rightarrow w_i\times e^{\alpha_mI(y_i\ne T_m(x_i))}.$$
Each classifier $T_m$ is required to be better than random guessing with 
respect to the weighted distribution upon which the classifier is trained.
Thus, $err_m$ is required to be less than 0.5, since, otherwise, the
weights would be updated in the wrong direction.
Next, renormalize the weights,
$w_i\rightarrow {w_i/ \sum_{i=1}^Nw_i}.$
The score for a given event is 
$$T(x) = \sum_{m=1}^{N_{tree}}\alpha_mT_m(x),$$
which is just the weighted sum of the scores over the individual trees, see Fig.2.

\begin{figure}
\epsfig{figure=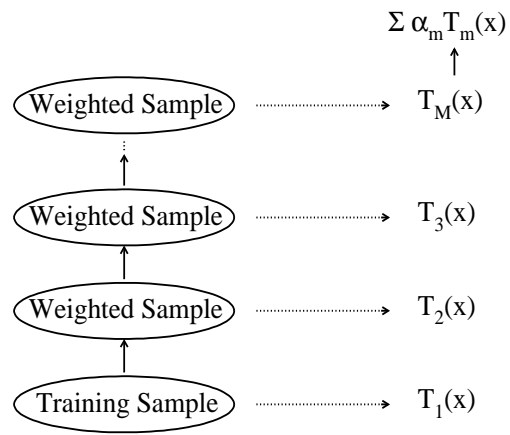,width=7cm,height=6cm}
\caption{Schematic of a boosting procedure.}
\end{figure}

The second boosting method is called
$\epsilon$-Boost\cite{eboost}, or sometimes ``shrinkage''.  
After the $m$th tree, change the weight of each event $i$, $i = 1,...,N$:
$$w_i\rightarrow w_ie^{2\epsilon I(y_i\ne T_m(x_i))},$$
where $\epsilon$ is a constant of the order of 0.01.
Renormalize the weights,
$w_i\rightarrow {w_i/ \sum_{i=1}^Nw_i}.$
The score for a given event is 
$$T(x)  = \sum_{m=1}^{N_{tree}}\epsilon T_m(x),$$
which is the renormalized, but unweighted, sum of the scores over individual
trees.

The AdaBoost and $\epsilon-$Boost algorithms used in this paper 
try to minimize the expectation
value: $E(e^{-yF(x)})$, where y = 1 for signal, y = -1 for background,
$F(x)=\sum_{i=1}^{N_{trees}} f_i(x)$, where the classifier $f_i(x)=1$ if an event lands
on signal leaf, and $f_i(x)=-1$ if an event lands on background leaf.
This minimization is closely related to minimizing the binomial
log-likelihood\cite{eboost}. It can be shown that $E(e^{-yF(x)})$ is minimized at
$$F(x)=\frac{1}{2}ln\frac{P(y=1|x)}{P(y=-1|x)}=\frac{1}{2}ln\frac{p(x)}{1-p(x)}$$
Let $y^* = (y+1)/2$. It is then easy to show that 
$$
e^{-yF(x)} = \frac{|y^*-p(x)|}{\sqrt{p(x)(1-p(x))}}
$$
The right-hand side is known as the $\chi$ statistic.
$\chi^2$ is a quadrative approximation to the log-likelihood, 
so $\chi$ can be considered a gentler alternative.
It turns out that fitting using $\chi$ is monotone and smooth;
the criteria will continually drive the estimates towards purer 
solutions. 
An ANN tries to minimize the squared-error 
$E(y-F(x))^2$, where y = 1 for signal events, y = 0 for background events,
and $F(x)$ is the network prediction for training events. 

\section{Results}

For the $\nu_\mu \rightarrow \nu_e$ oscillation search in the MiniBooNE experiment\cite{boone}, 
the main backgrounds come from  intrinsic $\nu_e$ contamination in the beam, 
mis-identified $\nu_\mu$ quasi-elastic scattering and mis-identified neutral 
current $\pi^0$ production.   Since intrinsic $\nu_e$ events are 
real $\nu_e$ events, the PID variables cannot distinguish them
from oscillation $\nu_e$ events.  This report concentrates on separating
the non-$\nu_e$ events from the $\nu_e$ events. Good sensitivity 
for the $\nu_e$ appearance search requires low background contamination from all kinds of
backgrounds. Here, the ANN and the 
two boosting algorithms are used to separate $\nu_e$ 
charged current quasi-elastic (CCQE) events from non-$\nu_e$ background 
events.

500000 Monte Carlo $\nu_\mu$ events distributed among the many 
possible final states and 200000 intrinsic $\nu_e$ 
CCQE events were fed into the reconstruction 
package R-fitter\cite{fitters}. 
Among these events, 88233 intrinsic $\nu_e$ CCQE and
162657 background events passed reconstruction and pre-selection cuts.
%(based on veto photomultiplier hit multiplicity, photomultiplier hit multiplicity, and fiducial volume).

The signature of each event is given by 52 variables for the R-fitter.
All variables are used in the boosting algorithms for training and testing.
It is a challenge to have agreement between data and Monte Carlo for
all of the PID variables and for the boosting outputs. The MiniBooNE
Collaboration is devoting considerable effort to achieve it.
Monte Carlo samples using 18 different parameter sets have been generated
and run through the same reconstruction programs. The results for both
the PID variables and the boosting outputs are consistent. 
When the present Monte Carlo is compared with the real data samples,
the shapes of the various PID variables and the boosting outputs match 
well. Since the recontruction and PID algorithms are still 
undergoing continuous modifications, relative results rather than 
absolute percentages are presented in the following plots.  

\begin{figure}
\epsfig{figure=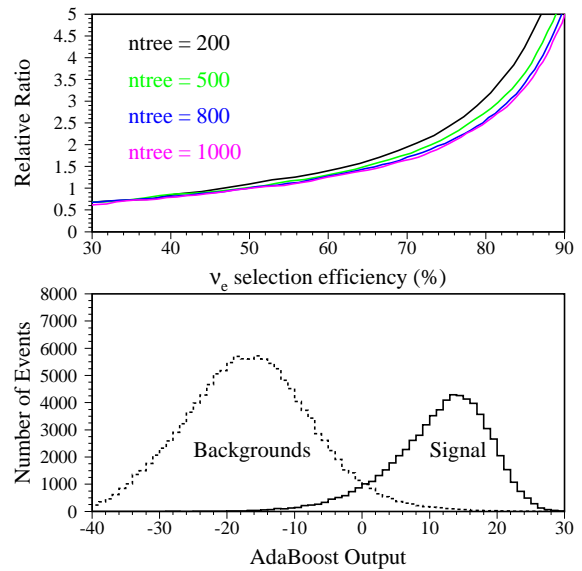,width=8cm}
\caption{Top: the number of background events kept divided by the number kept
for 50\% intrinsic $\nu_e$ selection efficiency and $N_{tree} = 1000$ versus 
the intrinsic $\nu_e$ CCQE selection efficiency.
Bottom: AdaBoost output, All kinds of backgrounds are combined for the boosting training.}
\end{figure}

For the AdaBoost algorithm, the parameter $\beta = 0.5$, the number of leaves 
$N_{leaves} = 45$ and the number of tree iterations $N_{tree} = 1000$ were used.
The relative ratio(defined as the number of background events kept divided by the 
number kept for 50\% intrinsic $\nu_e$ selection efficiency and $N_{tree} = 1000$)
as a function of $\nu_e$ selection efficiency for various tree iterations is shown in the 
top plot of Fig.3 and the AdaBoost output distributions are shown in the bottom plot.  
20000 intrinsic $\nu_e$ CCQE signal and 30000 background events were used for training,
68233  $\nu_e$ and 132657  background events were used for testing. All results shown
in the paper are for testing samples.

\begin{figure}
\epsfig{figure=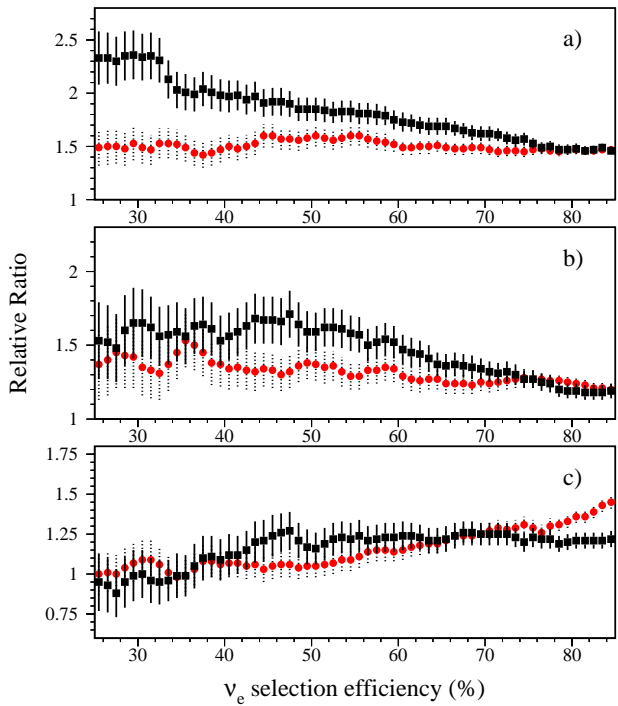,width=8.5cm}
\caption{Comparison of ANN and AdaBoost performance for test samples.
Relative ratio(defined as the number of background events kept for ANN
divided by the events kept for AdaBoost) versus the intrinsic $\nu_e$ CCQE 
selection efficiency.
a) all kinds of backgrounds are combined for the training against the signal.
b) trained by signal and neutral current $\pi^0$ background.
c) relative ratio is re-defined as the number of background events kept 
for AdaBoost with 21(red)/22(black) training variables divided by that for 
AdaBoost with 52 training variables.
All error bars shown in the figures are for Monte Carlo statistical errors only.
}
\end{figure}

In order to quantify the performance of the boosting algorithm,  
the AdaBoost results for a particular set of PID variables 
were compared with ANN results. The results, compared as a function of
the intrinsic $\nu_e$ CCQE selection efficiency, are shown in Fig.4.
For the intrinsic $\nu_e$ signal efficiency ranging from 40\% to 60\%,
the performances of AdaBoost were improved by a factor of approximately 
1.5 and 1.8 over the ANN if trained by the signal and 
all kinds of backgrounds with 21 (red dots) and 52 (black boxes) input 
variables respectively, shown in Fig.4.a.
If AdaBoost and ANN were trained by the signal and neutral current $\pi^0$ 
background, the performances of AdaBoost were improved by a factor of 
approximately 1.3 and 1.6 over the ANN for 22 (red dots) and 52 (black boxes) 
training variables respectively, shown in Fig.4.b. 
The best results for the ANN were found with 22 variables, 
while the best results for boosting were found with 52 variables. 
Comparison of the best ANN results and the best boosting results indicates 
that, when trained by the signal and neutral current $\pi^0$ background,
the ANN results kept approximately 1.5 times more background events than
were kept by the boosting algorithms for about 50\%  $\nu_e$ CCQE efficiencies.

In Fig.4.c, the ratio of the background kept for a 52 variable AdaBoost 
to that for a 21(red dots - results for AdaBoost trained
by the signal and all kinds of backgrounds) / 22(black boxes - results for 
AdaBoost trained by the signal and neutral current $\pi^0$ background)
variables is shown as a function of $\nu_e$ efficiency.
It can be seen that the AdaBoost performance is improved by the use of more
training variables.

The above ANN and AdaBoost performance comparison with different input variables 
indicates that AdaBoost can improve the PID performance significantly by using more
input variables, even though many of them have weak discriminant power; ANN, however,
seems unlikely to make full use of all input variables because it is 
more difficult to optimize all the weights between ANN nodes, given more nodes in 
both the input and the hidden layers. 
For the MiniBooNE Monte Carlo samples, the ANN are optimum for approximately
20 PID variables. The authors have found a similar number to be true for several
other applications. In general, the optimum number for ANN may 
vary depending on the strength of the PID variables and the correlations between them.

Further evidence of this effect comes from the S-fitter\cite{sfitters}, 
a second reconstruction--PID program set for the MiniBooNE.  A systematic
attempt was made to find the optimum sets of variables for ANN and for
boosting classifiers by using $\nu_e$ CCQE signal and $\pi^0$ background
(which includes 25 NUANCE reaction channels). It is found that, for
S-fitter, the optimum ANN result is achieved by a selected set of 22
variables, while for boosting, no obvious improvement is seen after a
selected optimum set of 50 variables are used.  Comparison of
the best ANN results and the best boosting results indicates that, for
a given fraction of $\nu_e$ CCQE events kept, the ANN results kept
about 1.2 times more $\pi^0$ background events than were kept by the
boosting algorithms within target range of keeping close to  50\% 
of the $\nu_e$ CCQE events.

\begin{figure}
\epsfig{figure=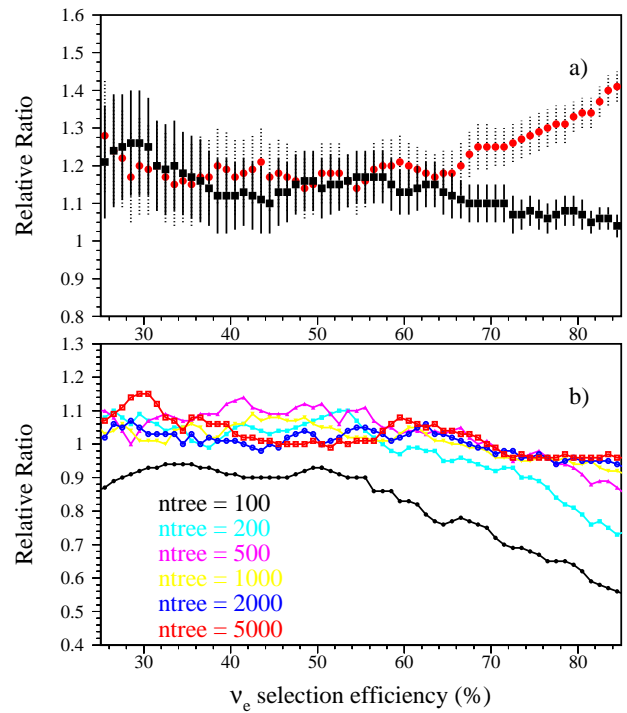,width=8.5cm}
\caption{Comparison of AdaBoost and $\epsilon$-Boost performance 
with different decision tree sizes (8 and 45 leaves per decision tree)
versus the intrinsic $\nu_e$ CCQE selection efficiency.
a) Relative ratio is defined as the number of background events kept 
for decision tree of 8 leaves divided by that for decision tree of 45 leaves,
red dots with error bars represent results from AdaBoost and black boxes
with error bars for $\epsilon$-Boost. The tree iterations
were 10000 for 8 leaves/tree and 1800 for 45 leaves/tree, respectively.
b) Relative ratio here is the number of background kept for AdaBoost 
divided by that for $\epsilon$-Boost with $N_{leaves} = 45$.  The
performance comparisons of AdaBoost and $\epsilon$-Boost with different
tree iterations are shown in different colors, $N_{tree} = $ 100(black), 200(cyan),
500(magenta), 1000(yellow), 2000(blue), 5000(red).}
\end{figure}

As noted in the introduction, two boosting algorithms are considered in the present paper.
The comparison of AdaBoost and $\epsilon$-Boost performance is shown in Fig.5, where
parameters $\beta=0.5$ and $\epsilon=0.01$ were selected for AdaBoost and $\epsilon$-Boost 
training, respectively. 
The comparison between small tree size (8 leaves) and large tree size (45 leaves) with 
a comparable overall number of decision leaves, indicates that large tree size with 45 leaves 
yields 10 $\sim$ 20 \% better performance for the MiniBooNE Monte Carlo samples
shown in Fig.5.a.  Increasing the tree size past 45 leaves did not produce
appreciable improvement

Comparison of AdaBoost and $\epsilon$-Boost performance for the background contamination 
versus the intrinsic $\nu_e$ CCQE selection efficiency as a function of the number of 
decision tree iterations is shown in Fig.5.b. A smaller relative ratio implies
a better performance for AdaBoost. The performance of AdaBoost is better than that of
$\epsilon$-Boost if the relative ratio is less than 1. Boosting performance in the  
high signal efficiency region is continuously improved for more tree iterations.
AdaBoost has better performance than $\epsilon$-Boost for less than about
200 tree iterations, but becomes slightly worse than $\epsilon$-Boost
for a large number of tree iterations, especially for $\nu_e$ signal 
efficiency below $\sim$ 60\%. For higher $\nu_e$ signal efficiency($>$ 70\%),
AdaBoost works slightly  better than  $\epsilon$-Boost.

%For a small decision tree size of 8 leaves, the performance of  
%$\epsilon$-Boost seems better than that of AdaBoost with tree iterations of 10000. 
%For a large decision tree size of 45 leaves with tree iterations of 1800,  
%$\epsilon$-Boost has slightly better performance than AdaBoost at low $\nu_e$ signal 
%efficiency($<$65\%), but worse at high $\nu_e$ signal efficiency($>$70\%). 

\section{Conclusions}

PID variables obtained 
using the R-fitter and the S-fitter event reconstruction programs 
for the MiniBooNE experiment were used to separate signal events from 
background events. The ANN and the boosting algorithms were compared 
for PID. Based on these studies with the MiniBooNE 
Monte Carlo samples, the boosting algorithms, AdaBoost  and $\epsilon$-Boost, 
improved PID performance significantly
compared with the artificial neural network technique. 
This improvement manifested itself when a large number of PID variables
was used.  For a small number of variables, the ANN classification was
competitive, but as the number of variables was increased, the boosting results
proved more efficient and superior to the ANN technique. If more 
variables are needed, boosting will use them as necessary.

It was also found that boosting with a large tree size of 45 leaves worked 
significantly better than boosting with a small tree size,
8 leaves, as recommended in some statistics literature.

The boosting technique proved to be quite
robust.  If a transformation of variables from $x$ to $y=f(x)$ is made,
then as long as the ordering is preserved, that is if $x_2>x_1$,
then $y_2>y_1$, the boosting results are unchanged.  ANNs must be tuned
for temperature, learning rate and other variables, while for boosting,
there is much less to vary and it is quite straightforward.

There are certainly applications where ANNs prove better than boosting.  However,
for this application boosting appears superior and seems to be exceptionally
robust and simple to use.  It is anticipated that boosting techniques will
have wide application in physics.

\section{Acknowledgments}
We wish to express our gratitude to the MiniBooNE Collaboration for their excellent
work on the Monte Carlo simulation and the software package for physics analysis.

This work is supported by the Department of Energy and by the 
National Science Foundation of the United States.

% Create the reference section using BibTeX:

%\bibliography{}

%%%%%%%%%%%%%%%%%%%%%%%% figures

%%%%%%%%%%%%%%%%%%%%%%%%%%
\end{document}